\documentclass{article}

\usepackage{latexsym}

 \newcommand{\sbs}{\subseteq} \newcommand{\Eqv}{\Longleftrightarrow}
 \newcommand{\eqv}{\Leftrightarrow} 
 \newcommand{\dam}{\Diamond} \newcommand{\sat}{{\models}}
 \newcommand{\dar}{{\downarrow}}
 \newcommand{\Ra}{\Rightarrow}
 \newcommand{\R}{{\bf R}}

 \newcommand{\sk}{{\sf K}}\newcommand{\sa}{{\sf A}}
 \newcommand{\sfl}{{\sf L}}
 \newcommand{\fsk}{{\hbox{\footnotesize\sk}}}

  \newcommand{\cl}{{\mathcal L}}
 \newcommand{\cb}{{\mathcal B}} \newcommand{\ct}{{\mathcal T}}
 \newcommand{\cm}{{\mathcal M}} \newcommand{\cp}{{\mathcal P}}
 \newcommand{\cf}{{\mathcal F}} \newcommand{\cg}{{\mathcal G}}
 \newcommand{\cn}{{\mathcal N}} \newcommand{\co}{{\mathcal O}}

 \newcommand{\Rem}[2]{{\sf Rem}^{#1}#2}
 
 \newcommand{\tsat}{\sat_{\ct}}
 \newcommand{\bsat}{\sat_{\cb}}
 \newcommand{\msat}{\sat_{\cm}}

\newtheorem{theorem}{\bf Theorem}
\newtheorem{lemma}[theorem]{\bf Lemma}
\newtheorem{proposition}[theorem]{\bf Proposition}
\newtheorem{corollary}[theorem]{\bf Corollary}

\newtheorem{claim}[theorem]{\bf Claim}

\newenvironment{definition}{\par\medskip\addtocounter{theorem}{1}%
  \noindent{\bf Definition \arabic{theorem}}\quad}{\medskip}

\newenvironment{example}{\par\medskip\addtocounter{theorem}{1}%
  \noindent{\bf Example \arabic{theorem}}\quad}{\medskip}

\newcommand{\qed}{\vrule height5pt width3pt depth0pt}

\newenvironment{proof}{\noindent {\it Proof.}}{{\nobreak\hfill \qed \par \medbreak}}

 \title{ Knowledge Theoretic Properties\\
                      of Topological Spaces}

 \author{Konstantinos Georgatos\\ Department of Mathematics\\
 Graduate School and University Center\\
 City University of New York\\
 33 West 42nd Street\\
 New York, NY 10036 }

\date{}

\begin{document}

 \maketitle

 \begin{abstract}
     We study the topological models of a logic of knowledge for
 topological reasoning, introduced by Larry Moss and Rohit Parikh (\cite{MP}).
     Among our results is a solution of a conjecture by
 the formentioned authors, finite satisfiability property and
 decidability for the theory of topological models.
 \end{abstract}

 \section{ Introduction}

      We are unable to measure natural  quantities
 with exact precision.  Physical devices or bounded resources always
 introduce a  certain amount  of error.  Because of  this fact, we are
 obliged to limit our observations to approximate values or, better,
 to sets of possible values.  Whenever this happens, sets of points,
 rather than points, are our subject of reasoning.
       Thus, the statement ``{\em  the  amount  of  ozone in the upper
 atmosphere has decreased by 12 per  cent\/}'' can never be known to
 be true with this precision.  What we mean is that e.g. the decrease
 has a value in the interval  $( 12-\epsilon,12+\epsilon )$ for  some
 positive real number $\epsilon$.  If we are able to spend more
 resources (taking more samples, using more precise  instruments, etc).
 we may be  able to affirm that the  value belongs to  a smaller
 interval  and therefore to refine our observation.  The topology of
 intervals in the real line  is our domain of reasoning.

     The above limitations do not prevent us from drawing  conclusions.
 In fact  it is  enough that  we {\em  know\/} that  a certain
 quantity belongs to  a set  of possible  values.  The  first
 hundred decimal points of $\pi$ are enough for most practical
 purposes and if we decide to settle for  such a value,  an algorithm
 that  computes these decimal points conveys the same knowledge as the
 actual algorithm that computes $\pi$.  What we know is exactly the
 common properties of all algorithms belonging to  the same  open set
 of the  algorithms we observe in the topology of  initial segments
 and this  notion of  knowledge coincides with   the   traditional
 one (Hintikka~\cite{HI}, Fagin et al.~\cite{FHV}, Halpern and Moses~\cite{HAM},
 Parikh and Ramanujam~\cite{PR}):
 what is known is whatever is true in all states  compatible with the observer's
 view.

   Increase  of  knowledge  is  strongly  linked  with  the amount of
 resources we  are willing  to invest.   An increase  of information
 is accompanied  by  an  increase  in  the  effort  of  acquiring it.
 This corresponds to the refinement of the open set in the relevant
 topology. In the formal system introduced in (Moss and Parikh~\cite{MP}) there
 are two basic modal operators; $\sk$ for {\em knowledge} and $\Box$ for
 {\em effort}.

    A  basic  advantage  of  this  logic  and its semantics over other
 temporal logics or logics of change is that, though we make no
 mention of  set,  we  are  able  to  interpret  assertions relative to
 a set of possible  states  and,  at  the  same  time, keep the
 dependence on the actual state.  Topology is a  tool for modelling
 classes of  statements with  an  intuitionistic  flavor  such  as
 refutative  or  affirmative assertions (see (Vickers~\cite{V})) and this logic
 system enables us to treat them in a classical modal framework.
 In many respects the way we interpret the modal operator $\sk$
 resembles the algebraic semantics of a modal operator used to interpret
 intuitionistic statements as in (Rasiowa and Sikorski~\cite{RS}). As the
 intuitionistic
 version of a statement is the interior of the subset that represents
 the classical version of it, $\sk A$ is satisfied only in the open
 subsets  which are subsets of the interior of the set of points which
 satisfy a property $A$.

  The fundamental reasoning that this logic tries
 to capture has many equivalents in recursion theory and elsewhere in
 Mathematics. The discussion of them is  well beyond the scope  of this
 paper and the reader is referred to (Georgatos~\cite{KGT}) and (Moss
 and Parikh~\cite{MP}) for a more detailed exposition.

    In the following section, we  describe the syntax and semantics  of
 the logic and we give  complete axiomatisations with respect to
 subset spaces and topological spaces.  In section 3 we  develop a theory for
 describing the validity problem in topological spaces.  In  section 4 we study
 the model based on the basis of  a topological space closed under finite
 unions, and we prove it equivalent to the topological space that it
 generates. These results translate to a completeness theorem for topologies,
 given a finite axiomatisation for the class of spaces which are closed
 under (finite) intersection and union. In the last section we prove finite
 satisfiability for the class of topological models and decidability for their
 theory.

 \section{The logic}

 We follow the notation of (Moss
 and Parikh~\cite{MP}).

 Our language is bimodal and propositional.
 Formally, we start with a countable set $\sa$ of {\em atomic
 formulae} containing two distinguished elements $\top$ and $\bot$.
 Then the {\em language} $\cl$ is the
 least set such that $\sa\sbs\cl$ and closed under the following rules:

 $$\frac{\phi,\psi\in\cl}{\phi\land\psi\in\cl}\qquad
 \frac{\phi\in\cl}{\neg\phi,\Box\phi,\sk\phi\in\cl}$$

 \medskip
 The above language can be interpreted inside any spatial context.

 \begin{definition}
 Let $X$ be a set and $\co$ a subset of the powerset of $X$, i.e.
 $\co\sbs\cp(X)$ such that $X\in\co$.
 We call the pair $\langle X,\co \/\rangle$ a {\em subset space}.
 A {\em model} is a triple $\langle X,\co,i\/\rangle$, where
 $\langle X,\co \rangle$ is a subset space and $i$ a map from $\sa$
 to $\cp(X)$ with $i(\top)=X$ and $i(\bot)=\emptyset$ called {\em
 initial interpretation}.
 \end{definition}

 We denote the set
 $\{ (x,U) \mid   U\in\co,\  x\in U\}\sbs X\times\co$
 with $X\dot{\times}\co$. For each $U\in\co$ let $\dar U$ be the set
 $\{ V \mid  V\in\co \hbox{ and } V\sbs U \}$, i.e. the lower closed set
 generated by $U$ in the partial order $(\co,\sbs)$.

 \begin{definition}
 The {\em satisfaction relation} $\msat$, where $\cm$ is the model
 $\langle X,\co,i\/\rangle$,
 is a subset of $(X\dot{\times}\co)\times\cl$ defined recursively  by
 (we write $x,U\msat\phi$ instead of $((x,U),\phi)\in\msat$):
 $$
 \begin{array}{ll}
 x,U\msat A & \hbox{iff}\quad x\in i(A),\hbox{ where }A\in\sa \medskip
 \\ x,U\msat \phi\land\psi & \hbox{if}\quad x,U\msat\phi \hbox{ and }
                                            x,U\msat \psi\medskip  \\
 x,U\msat \neg\phi & \hbox{if}\quad x,U \not\msat\phi        \medskip
 \\ x,U\msat \sk\phi & \hbox{if}\quad\hbox{for all } y\in U,\quad
                                            y,U\msat\phi \medskip  \\
 x,U\msat \Box\phi & \hbox{if} \quad\hbox{for all } V\in\dar U \hbox{
 such that } x\in V, \quad x,V\msat\phi.
 \end{array}
 $$
 If $x,U\msat\phi$ for all $(x,U)$ belonging to $ X\dot{\times}\co$
 then $\phi$ is {\em valid} in $\cm$, denoted by $\cm\sat\phi$.
 \end{definition}

 We abbreviate $\neg\Box\neg\phi$ and $\neg\sk\neg\phi$ with
 $\dam\phi$ and $\sfl\phi$ respectively.
 We have that
 $$
 \begin{array}{ll}
 x,U\msat \sfl\phi & \hbox{if there exists } y\in U \hbox{ such that }
                                         y,U\msat\phi\ \hbox{and},\medskip\\
 x,U\msat \dam\phi & \hbox{if there exists } V\in\co \hbox{ such that }
 V\sbs U,\  x\in V,\hbox { and } x,V\msat\phi.
 \end{array}
 $$

 Many  topological properties  are expressible in this logical
 system in a natural way. For instance, in a model where the subset
 space is a topological space,  $i(A)$ is  {\em open} whenever
 $A\to\dam\sk A$ is valid in this model or $i(A)$ is {\em nowhere
 dense} whenever $\sfl\dam\sk\neg A$ is valid (cf. (Moss
 and Parikh~\cite{MP})).

 \begin{example}
 Consider the set of {\em real numbers} $\R$ with the usual
 topology of open intervals. We define the following three
 predicates:

 \begin{eqnarray*}
 {\tt pi} & \hbox{where} & i({\tt pi}) = \{ \pi \} \\
 {\tt I}_1 & \hbox{where} & i({\tt I}_1)=(-\infty,\pi ]\\
 {\tt I}_2 & \hbox{where}  & i({\tt I}_2)=(\pi,+\infty)\\
 {\tt Q} & \hbox{where} & i({\tt Q})=\{ q \mid  q\hbox{ is rational }\}.
 \end{eqnarray*}

 There is no real number $p$ and open set $U$ such that
 $p,U\sat\sk {\tt pi}$ because that would imply $p=\pi$ and
 $U=\{\pi\}$ and there are no singletons which are open.

 A point $x$ belongs to the {\em closure} of a set $W$ if every
 open $U$ that contains $x$ intersects $W$.
 Thus $\pi$ belongs to the closure of $(\pi,+\infty)$, i.e every open
 that contains $\pi$ has a point in $(\pi,+\infty)$. This means that
 for all $U$ such that $\pi\in U$, $\pi,U\sat\sfl {\tt I}_2$,
 therefore $\pi,\R\sat\Box\sfl {\tt I}_2$. Following
 the same reasoning $\pi,\R \sat\Box \sfl {\tt I}_1$, since $\pi$
 belongs to the closure of $(-\infty,\pi ]$.

 A point $x$ belongs to the {\em boundary} of  a set $W$ whenever
 $x$ belong to the closure of $W$ and $X-W$. By the above, $\pi$
 belongs to the boundary of $(-\infty,\pi ]$ and
 $\pi,\R \sat \Box ( \sfl {\tt I}_1 \land \sfl {\tt I}_2 )$.

 A set $W$ is {\em closed} if it contains its closure. The interval
 $i({\tt I}_1)=(-\infty,\pi ]$ is closed and this means that the
 formula $\Box\sfl {\tt I}_1 \to {\tt I}_1$ is valid.

 A set $W$ is {\em dense} if all opens contain a point of $W$. The set
 of rational numbers is dense which translates to the fact that the
 formula $\Box\sfl {\tt Q}$ is valid. To exhibit the reasoning in this
 logic, suppose that the set of rational numbers was closed then both
 $\Box\sfl {\tt Q}$ and $\Box\sfl {\tt Q} \to {\tt Q}$ would be valid.
 This implies that ${\tt Q}$  would be valid which means that all reals
 would be rationals. Hence the set of rational numbers is not closed.
 \end{example}

 The following set of axioms and rules, denoted by $\bf MP^*$, is sound and
 complete for the class of topological spaces (see (Georgatos~\cite{KGT})) while
 axioms 1
 through~\ref{ax:boxk}, denoted by $\bf MP$, appeared first and proven sound and
 complete for the class of subset spaces in (Moss
 and Parikh~\cite{MP}).
 \medskip

 \noindent {\bf Axioms}
 \begin{enumerate}
 \item All propositional tautologies
 \item $(A\to\Box A) \land (\neg A\to\Box\neg A)$, for $A\in\sa$
 \item $\Box(\phi\to\psi)\to(\Box\phi\to\Box\psi)$
 \item $\Box\phi\to\phi$
 \item $\Box\phi\to\Box\Box\phi$
 \item $\sk(\phi\to\psi)\to(\sk\phi\to\sk\psi)$
 \item $\sk\phi\to\phi$
 \item $\sk\phi\to\sk\sk\phi$
 \item $\phi\to\sk\sfl\phi$
 \item $\sk\Box\phi\to\Box\sk\phi$\label{ax:boxk}
 \item $\dam\Box\phi\to\Box\dam\phi$\label{ax:direct}
 \item $\dam(\sk\phi\land\psi)
        \land\sfl\dam(\sk\phi\land\chi)
        \to\dam(\sk\dam\phi\land\dam\psi\land\sfl\dam\chi)$\label{ax:union}
 \end{enumerate}
 \pagebreak[2]
 \noindent{\bf Rules}
 $$\frac{\phi\to\psi,\phi}{\psi}\ \hbox{\footnotesize MP}$$
 $$\frac{\phi}{\sk\phi}\ \hbox{\footnotesize
 \footnotesize \sk-Necessitation} \qquad
   \frac{\phi}{\Box\phi}\ \hbox{\footnotesize $\Box$-Necessitation}$$
 \medskip
 \pagebreak[2]

 \section{Stability and Splittings}
 \label{sec:stability}
 Suppose that $X$ is a set and $\ct$ a topology on $X$.  In the
 following we assume that we are working in the topological space
 $( X,\ct )$. Our aim  is to find a partition of $\ct$, where a given
 formula $\phi$ ``retains its truth value'' for each point throughout a
 member of this partition. We shall show that there exists a finite
 partition of this kind.

 \begin{definition}
 Given a finite family $\cf=\{ U_1,\ldots,U_n\}$ of opens, we define
 the {\em remainder} of (the principal ideal in $(\ct,\sbs)$ generated
 by) $U_k$ by
 $$\Rem{\cf}{U_k}\quad=\quad\dar U_k - \bigcup_{U_k\not\subseteq U_i}\dar U_i.$$
 \end{definition}

 \begin{proposition} In a finite set of opens $\cf=\{U_1,\ldots,U_n\}$ closed
 under intersection, we have
 $$\Rem{\cf}{U_i}\quad=\quad\dar U_i - \bigcup_{U_j\subset U_i}\dar U_j,$$
 for $i=1,\ldots,n$. \end{proposition}

 \begin{proof}
 $$
 \begin{array}{rcl}
 \Rem{\cf}{U_i} & = & \dar U_i - \bigcup_{U_i\not\subseteq U_h}\dar U_h \\
              & = & \dar U_i - \bigcup_{U_i\not\subseteq U_h}\dar (U_h\cap U_i)\\
              & = & \dar U_i - \bigcup_{U_j\subset U_i}\dar U_i. \\
 \end{array}
 $$
 \end{proof}

 We denote $\bigcup_{U_i\in\cf}\dar U_i$ with $\dar\cf$.

 \begin{proposition} If $\cf=\{U_1,\ldots,U_n\}$ is a finite family of opens,
 closed under intersection, then
 \label{prop:part}
 \renewcommand{\theenumi}{\alph{enumi}}
 \begin{enumerate}
 \item
 $\Rem{\cf}{U_i}\cap\Rem{\cf}{U_j}=\emptyset$, for $i\not= j$,
 \item
 $\bigcup^n_{i=1}\Rem{\cf}{U_i}=\dar\cf$,
 i.e.
 $\{\Rem{\cf}{U_i}\}^n_{i=1}$ is a partition of $\dar\cf$. We call
 such an $\cf$ a {\em finite splitting (of $\dar\cf$\/)\/},
 \item
 if $V_1,V_3\in\Rem{\cf}{U_i}$ and $V_2$ is an open such that
 $V_1\sbs V_2\sbs V_3$ then $V_2\in\Rem{\cf}{U_i}$, i.e.
 $\Rem{\cf}{U_i}$ is convex.\label{convex}
 \end{enumerate}
 \end{proposition}

 \begin{proof}
 The first and the third are immediate from the definition.

 For the second, suppose that $V\in\dar\cf$ then
 $V\in\Rem{\cf}{\bigcap_{V\in\dar U_i}U_i}$.
 \end{proof}

 Every partition of a set induces an equivalence relation on this set.
 The members of the partition comprise the equivalence classes. Since a
 splitting induces a partition, we denote the equivalence relation induced by a
 splitting $\cf$ by $\sim_\cf$.

 \begin{definition}
 Given a set of open subsets $\cg$, we define the relation $\sim'_\cg$
 on $\ct$ with $V_1\sim'_\cg V_2$ if and only if $V_1\sbs U\eqv V_2\sbs
 U$ for all $U\in\cg$.
 \end{definition}

 We have the following

 \begin{proposition}
 The relation $\sim'_\cg$ is an equivalence.
 \end{proposition}

 \begin{proposition}
 Given a finite splitting $\cf$, $\sim'_\cf = \sim_\cf$ i.e. the
 remainders of $\cf$ are the equivalence classes of $\sim'_\cf$.
 \label{prop:equiv}\end{proposition}

 \begin{proof}
 Suppose $V_1\sim'_\cf V_2$ then $V_1,V_2\in \Rem{\cf}{U}$, where
 $$U\quad=\quad\bigcap\{\ U'\ \mid V_1,V_2\sbs U,\ U'\in\cf\ \}.$$
 For the other way suppose $V_1,V_2\in\Rem{\cf}{U}$ and that there
 exists $U'\in\cf$ such that $V_1\sbs U'$ while $V_2\not\sbs U'$.
 Then we have that $V_1\sbs U'\cap U$, $U'\cap U\in\cf$ and
 $U'\cap U\subseteq U$ i.e. $V_1\not\in\Rem{\cf}{U}$.
 \end{proof}

 We state some useful facts about splittings.

 \begin{proposition}
 If $\cg$ is a finite set of opens, then ${\sf Cl(\cg)}$,
 its closure under intersection, yields a finite splitting for $\dar\cg$.
 \label{prop:closinter}
 \end{proposition}

 The last proposition enables us to give yet another characterization
 of remainders: every family of points in a complete lattice
 closed under arbitrary joins comprises a
 {\em closure system}, i.e. a set of fixed points of a closure operator
 of the lattice (cf. (Gierz et al.~\cite{COMP})). Here, the lattice is the poset
 of the opens of the topological space. If we restrict ourselves to a finite
 number of fixed points then we just ask for a finite set of opens
 closed under intersection i.e. Proposition~\ref{prop:closinter}.
 Thus a closure operator in the lattice of the open subsets of a
 topological space induces an equivalence relation, two opens being
 equivalent if they have the same closure, and the equivalence
 classes of this relation are just the remainders of the open subsets
 which are fixed points  of the closure operator.
 The maximum open in $\Rem{\cf}{U}$, i.e. $U$, can be taken as the representative
 of the equivalence class which is the union of all open
 sets belonging to $\Rem{\cf}{U}$.

 We now introduce the notion of stability corresponding to what we
 mean by ``a formula retains its truth value on a set of opens''.

 \begin{definition}
 If $\cg$ is a set of opens then $\cg$ is {\em stable for
 $\phi$}, if for all $x$, either $x,V\sat\phi$ for all $V\in\cg$,
 or $x,V\sat\neg\phi$ for all $V\in\cg$, such that $x\in V$.
 \end{definition}

 \begin{proposition}
 If $\cg_1$,$\cg_2$ are sets of opens then
 \label{prop:rem}
 \renewcommand{\theenumi}{\alph{enumi}}
 \begin{enumerate}
 \item if $\cg_1\sbs\cg_2$ and $\cg_2$ is stable for
 $\phi$ then $\cg_1$ is stable for $\phi$ ,\label{subrem}
 \item
 if $\cg_1$ is stable for $\phi$ and $\cg$ is stable for $\chi$ then
 $\cg_1\cap\cg_2$ is stable for $\phi\land\chi$.\label{interem}
 \end{enumerate}
 \end{proposition}

 \begin{proof} (\ref{subrem}) is easy to see while (\ref{interem}) is a
 corollary of (\ref{subrem}).
 \end{proof}

 \begin{definition}
 A finite splitting $\cf=\{U_1,\ldots,U_n\}$ is called
 a {\em stable splitting for} $\phi$, if $\Rem{\cf}{U_i}$ is stable
 for $\phi$ for all $U_i\in\cf$.
 \end{definition}

 \begin{proposition} If $\cf=\{U_1,\ldots,U_n\}$ is a stable splitting for
 $\phi$, so is
 $${\cf}'={\sf Cl}(\{U_0,U_1,\ldots,U_n\}),$$
 where $U_0\in\dar\cf$.\label{prop:unfsp}  \end{proposition}

 \begin{proof}
 Let $V\in\cf'$ then there exists $U_l\in\cf$ such that
 $\Rem{\cf'}{V}\sbs\Rem{\cf}{U_l}$ (e.g.
 $U_l=\bigcap\{U_i\mid U_i\in\cf,V\sbs U_i\}$) i.e. $\cf'$ is a
 {\em refinement} of $\cf$. But $\Rem{\cf}{U_l}$ is stable for $\phi$
 and so is $\Rem{\cf'}{V}$
 by Proposition~{\ref{prop:rem}(\ref{subrem})}.
 \end{proof}

 The above proposition tells us that if there is a
 finite stable splitting for a topology then there is a closure
 operator with finitely many fixed points whose associated equivalence classes
 are stable sets of open subsets.

 Suppose that  $\cm=\langle X,\ct,i\rangle$ is a topological model for
 $\cl$. Let $\cf_\cm$ be a family of subsets of $X$ generated as follows:
 $i(A)\in\cf_\cm$ for all $A\in\sa$, if $S\in\cf_\cm$ then
 $X-S\in\cf_\cm$, if $S,T\in\cf_\cm$ then $S\cap T\in\cf_\cm$, and if
 $S\in\cf_\cm$ then $S^\circ\in\cf_\cm$ i.e. $\cf_\cm$ is the least set
 containing $\{i(A)\mid A\in\sa\}$ and closed under complements,
 intersections and interiors. Let $\cf^\circ_\cm$ be the set
 $\{S^\circ\mid S\in\cf_\cm\}$. We have
 $\cf^\circ_\cm=\cf_\cm\cap\ct$. The following is the main theorem of this
 section.

 \begin{theorem}[Partition Theorem]
 Let $\cm=\langle X,\ct,i\rangle$ be a topological model. Then there exists
 a set $\{\cf^\psi\}_{\psi\in\cl}$
 of finite stable splittings such that
 \begin{enumerate}
 \item $\cf^{\psi}\sbs\cf^\circ_\cm$ and $X\in\cf^{\psi}$,  for all $\psi\in\cl$,
 \item if $U\in\cf^\psi$ then $U^\psi=\{x\in U\mid x,U\sat\psi\}\in\cf_\cm$, and
 \item if $\phi$ is a
        subformula of $\psi$ then $\cf^{\phi}\sbs\cf^{\psi}$ and $\cf^{\psi}$
        is a finite stable splitting for $\phi$,
 \end{enumerate}
 where $\cf_\cm$, $\cf_\cm^\circ$ as above.
 \label{thm:main}
 \end{theorem}

 \begin{proof}
 By induction on the structure of the formula $\psi$. In
 each step we take care to refine the partition of the induction
 hypothesis.

 \begin{itemize}

 \item If $\psi=A$ is an atomic formula, then
 $\cf^A=\{X,\emptyset\}=\{i(\top),i(\bot)\}$, since $\ct$ is stable for
 all atomic formulae. We also have $\cf^A\sbs\cf^\circ_\cm$ and
 $X^A=i(A)\in\cf_\cm$.

 \item If $\psi=\neg\phi$ then let $\cf^{\psi}=\cf^{\phi}$,
 since the statement of the proposition is symmetric with respect to
 negation. We also have that for an arbitrary $U\in\cf^\psi$,
 $U^\psi=U^\neg\phi$.

 \item If $\psi=\chi\land\phi$, let
 $$\cf^{\psi}={\sf Cl}(\cf^{\chi}\cup\cf^{\phi}).$$
 Observe that
 $\cf^{\chi}\cup\cf^{\phi}\sbs\cf^{\chi\land\phi}$.

 Now, if $W_i\in\cf^{\psi}$ then there exists
 $U_j\in\cf^{\chi}$ and $V_k\in\cf^{\phi}$ such that
 $$W_i=U_j\cap V_k \quad\hbox{and}\quad
 \Rem{\cf^{\psi}}{W_i}\sbs\Rem{\cf^{\chi}}{U_
 j } \cap\Rem{\cf^{\phi}}{V_k}$$
 (e.g. $U_j=\bigcap\{U_m\mid W_i\sbs U_m,\ U_m\in\cf^{\chi}\}$ and
       $V_k=\bigcap\{V_n\mid W_i\sbs V_n,\ V_n\in\cf^{\phi}\}$).
 Since $\Rem{\cf^{\chi}}{U_j}$ is stable for $\chi$ and
 $\Rem{\cf^{\phi}}{V_n}$ is stable for $\phi$, their intersection
 is stable for $\chi\land\phi=\psi$, by
 Proposition~{\ref{prop:rem}(\ref{interem})}, and so is its
 subset $\Rem{\cf^{\psi}}{W_i}$, by
 Proposition~{\ref{prop:rem}(\ref{subrem})}.
 Thus $\cf^{\psi}$ is a finite stable splitting for $\psi$
 containing $X$.

 We have that $\cf^\psi\sbs\cf_\cm$ whenever $\cf^\chi\sbs\cf_\cm$ and
 $\cf^\phi\sbs\cf^\circ_\cm$. Finally, $W^\psi_i=U^\chi_j\cap
 V^\phi_k$.

 \item Suppose $\psi=\sk\phi$. Then, by induction
 hypothesis, there exists a finite stable
 splitting $\cf^{\phi}=\{U_1,\ldots,U_n\}$
 for $\phi$ containing $X$. Let $$W_i=(U^\phi_i)^{\circ},$$
 for all $i\in\{1,\ldots,n\}$.

 Observe that if $x\in U_i-W_i$ then $x,V\sat\neg\phi$, for all
 $V\in\Rem{\cf^{\phi}}{U_i}$ and $x\in V$, since
 $\Rem{\cf^{\phi}}{U_i}$ is stable for $\phi$, by induction hypothesis.

 Now, if $V\in\Rem{\cf^{\phi}}{U_i}\cap\dar W_i$, for some
 $i\in\{1.\ldots,n\}$, then $x,V\sat\phi$ for
 all $x\in V$, by definition of $W_i$, hence $x,V\sat\sk\phi$ for all
 $x\in V$.

 On the other hand, if $V\in\Rem{\cf^{\phi}}{U_i}-\dar W_i$ then there
 exists $x\in V$ such that $x,V\sat\neg\phi$ (otherwise $V\sbs W_i$ ).
 Thus we have $x,V\sat\neg\sk\phi$ for all $x\in V$.
 Hence $\Rem{\cf^{\phi}}{U_i}\cap\dar W_i$ and
 $\Rem{\cf^{\phi}}{U_i}-\dar W_i$ are stable for $\sk\phi$.
 Thus, the set
 $$F=\{\Rem{\cf}{U_i}\mid W_i\not\in\Rem{\cf}{U_i}\}
 \cup \{\Rem{\cf}{U_j}-\dar W_j,\Rem{\cf}{U_j}\cap\dar W_j\mid W_j\in
 U_j\}$$
 is a partition of $\ct$ and its members are stable for $\sk\phi$.
 Let $\sim_F$ be the equivalence relation on $\ct$ induced by $F$ and
 let
 $$\cf^{\fsk\phi}={\sf Cl}(\cf^{\phi}\cup\{\ W_i\ \mid \
 W_i\in\Rem{\cf^{\phi}}{U_i}\}).$$
 We have that $\cf^{\fsk\phi}$ is a finite set of opens and
 $\cf^{\phi}\sbs\cf^{\fsk\phi}$. Thus, $\cf^{\fsk\phi}$ is  finite
 and contains $X$. We have only to prove that $\cf^{\fsk\phi}$
 is a stable splitting for $\sk\phi$, i.e. every remainder of an open
 in $\cf^{\fsk\phi}$ is stable for $\sk\phi$.

 If $V_1\not\sim_F V_2$, where $V_1,V_2\in\ct$, then there exists
 $U=U_i$ or $W_i$ for some $i=1,\ldots,n$ such that $V_1\sbs U$ while
 $V_2\not\sbs U$. But this implies that $V_1\not\sim_{\cf^{\fsk\phi}}
 V_2$. Therefore $\{\Rem{\cf^{\fsk\phi}}{U}\}_{U\in\cf^{\fsk\phi}}$ is a
 refinement of $F$ and $\cf^{\fsk\phi}$ is a finite stable splitting
 for $\sk\phi$ using Proposition~{\ref{prop:rem}(\ref{subrem})}.

 We have that $\cf^{\fsk\phi}\sbs\cf^\circ_\cm$ because
 $W_i\in\cf^\circ_\cm$, for $i=1,\ldots,n$. Now if $U\in\cf^\psi$ then
 either $U^{\sk\phi}=U$ or $U^{\sk\phi}=\emptyset$.

 \item Suppose $\psi=\Box\phi$. Then, by induction
 hypothesis, there exists a finite stable splitting
 $\cf^{\phi}=\{U_1,\ldots,U_n\}$ for $\phi$ containing $X$.

 Let
 $$\cf^{\Box\phi}={\sf Cl}(\cf^{\phi}\cup\{U_i\Ra U_j \mid 1\leq
 i,j\leq n\}),$$
 where $\Ra $ is the implication of the complete Heyting algebra $\ct$
 i.e. $V\sbs U\Ra W$ if and only if $V\cap U\sbs W$ for $V,U,W\in\ct$.
 We have that $U\Ra W$ equals $(X-(U-W))^\circ$. Clearly,
 $\cf^{\Box\phi}$ is a finite splitting containing $X$ and
 $\cf^{\phi}\sbs\cf^{\Box\phi}$. We have only to prove that
 $\cf^{\Box\phi}$ is stable for $\Box\phi$. But first, we prove the
 following  claim:

 \begin{claim}
 Suppose $U\in\cf^{\phi}$ and $U'\in\cf^{\Box\phi}$. Then
 $$U'\cap U\in\Rem{\cf^{\phi}}{U}\quad\Eqv\quad
 V\cap U\in\Rem{\cf^{\phi}}{U}\hbox{ for all }
 V\in\Rem{\cf^{\Box\phi}}{U'}.$$
 \label{claim2}
 \end{claim}

 \begin{proof}
 The one direction is straightforward. For the other, let
 $V\in\Rem{\cf^{\Box\phi}}{U'}$ and suppose
 $V\cap U\not\in\Rem{\cf^{\phi}}{U}$ towards a contradiction.
 This implies that there exists $U''\in\cf^{\phi}$, with $U''\subset
 U$, such that $V\cap U\sbs U''$. Thus, $V\sbs U\Ra U''$ but
 $U'\not\sbs U\Ra U''$. But $U\Ra U''\in\cf^{\Box\phi}$ which
 contradicts $U'\sim_{\cf^{\Box\phi}} V$, by
 Proposition~{\ref{prop:equiv}}.
 \end{proof}

 Let $U'\in\cf^{\Box\phi}$. We must prove that
 $\Rem{\cf^{\Box\phi}}{U'}$ is stable for $\Box\phi$.

 Suppose that $x,U'\sat\neg\Box\phi$. We must prove that
 $$x,V'\sat\neg\Box\phi$$
 for all $V'\in\Rem{\cf^{\Box\phi}}{U'}$ such
 that $x\in V'$.

 Since $x,U'\sat\neg\Box\phi$, there exists $V\in\ct$,
 with $x\in V$ and $V\sbs U'$, such that $x,V\sat\neg\phi$.
 Since $\cf^{\phi}$ is a splitting, there exists $U\in\cf^{\phi}$ such
 that $V\in\Rem{\cf^{\phi}}{U}$. Observe that $V\sbs U'\cap U\sbs U$,
 so $U'\cap U\in\Rem{\cf^{\phi}}{U}$, by
 Proposition~{\ref{prop:part}(\ref{convex})}.

 By Claim~\ref{claim2}, for all $V'\in\Rem{\cf^{\Box\phi}}{U'}$,
 we have $V'\cap U\in\Rem{\cf^{\phi}}{U}$.
 Thus if $x\in V'$ then $x,V'\cap U\sat\neg\phi$, because
 $\Rem{\cf^{\phi}}{U}$ is stable for $\phi$ by induction hypothesis.
 This implies that, for all $V'$ such that
 $V'\in\Rem{\cf^{\Box\phi}}{U'}$ and $x\in V$, we have
 $x,V'\sat\neg\Box\phi$.

 Therefore, $\cf^{\Box\phi}$ is a finite stable splitting for
 $\Box\phi$.

 Now $U_i\Ra U_j\in\cf^\circ_\cm$ for $1\leq i,j\leq n$,  hence
 $\cf^{\Box\phi}\sbs\cf^\circ_\cm$.

 Finally, let $U$ belong to $\cf^{\Box\phi}$ and
 $V_1,\ldots,V_m$ be all opens in $\cf^\phi$ such that
 $U\cap V_i\in\Rem{\cf^\phi}{V_i}$, for $i=1,\ldots,m$.
 Then $x,U\sat\dam\neg\phi$ if and only if there exists
 $j\in\{1,\ldots,m\}$ with $x\in V_j$ and $x,V_j\sat\neg\phi$ because
 $x,V_j\cap U\sat\neg\phi$ since $V_j\cap U\in\Rem{\cf^\phi}{V_j}$.
 This implies that
 $$U^{\neg\Box\phi}\quad=\quad
 U^{\dam\neg\phi}\quad=\quad U\cap\bigcup^m_{i=1} V^{\neg\phi}_i.$$
 Since $U,V^{\neg\phi}_1,\ldots,V^{\neg\phi}_m$ belong to $\cf_\cm$, so
 does $U^{\neg\Box\phi}$ and, therefore,
 $U^{\Box\phi}=U-U^{\neg\Box\phi}$.

 \end{itemize}

 In all steps of induction we refine the finite splitting, so
 if $\phi$ is a subformula of $\psi$ then
 $\cf^{\phi}\sbs\cf^{\psi}$ and $\cf^{\psi}$ is stable for
 $\phi$ using Proposition~{\ref{prop:rem}(\ref{subrem})}.
 \end{proof}

 Theorem~\ref{thm:main} gives us a great deal of intuition for
 topological models. It describes in detail the expressible
 part of the topological lattice for the completeness result as it
 appears in (Georgatos~\cite{KGT}) and paves the road for
 the reduction of the theory of topological models to that of spatial
 lattices and the decidability result of this section.


 \section{Basis Model}

      Let $\ct$ be a topology on a set $X$ and $\cb$ a basis for $\ct$.
 We denote satisfaction in the models $\langle X,\ct,i\/\rangle$ and
 $\langle X,\cb,i\/\rangle$ by $\tsat$ and $\bsat$, respectively.
 In the following proposition we prove that each equivalence class
 under $\sim_{\cf}$ contains an element of  a basis closed under finite
 unions.

 \begin{proposition}
      Let $(X,\ct)$ be a topological space, and let $\cb$ be a basis
 for $\ct$ closed under finite unions. Let $\cf$ be any finite subset
 of $\ct$. Then for all $V \in \cf $ and all $x\in V$, there is some $U
 \in \cb$ with $x \in U \subseteq V$ and $U\in\Rem{\cf}{V}$.
 \label{prop:basis}
 \end{proposition}

 \begin{proof}
      By finiteness of $\cf$, let $V_1, \ldots, V_k$ be the  elements
 of $\cf$ such that $V\not\sbs V_i$, for $i\in\{1,\ldots,k\}$. Since
 $V_i\not=V$, take $x_i\in V-V_i$ for $i\in\{1,\ldots,k\}$. Since $\cb$
 is a basis for $\ct$, there exist $U_x,U_i$, with $x\in U_x$ and
 $x_i\in U_i$, such that $U_x$ and $U_i$ are subsets of $V$ for
 $i\in\{1,\ldots,k\}$. Set
 $$U=(\bigcup^k_{i=1} U_i)\cup U_x.$$
 Observe that $x\in U$, and $U\in\cb$, as it is a finite union of
 members of $\cb$. Also $U\in\Rem{\cf}{V}$, since $U\in\dar V$ but
 $U\not\in\bigcup\dar V_i$ for $i\in\{1,\ldots,k\}$.
 \end{proof}

 \begin{corollary}
      Let $(X,\ct)$ be a topological space, $\cb$ a basis for $\ct$
 closed under finite unions, $x\in X$ and $U\in\cb$. Then
 $$x,U\tsat\phi\quad\Eqv\quad x,U\bsat\phi.$$
 \label{corol:subsatm}
 \end{corollary}

 \begin{proof}
 By induction on $\phi$.

       The interesting case is when $\phi = \Box\psi$. Fix $x$, $U$,
 and $\psi$. By Proposition~\ref{thm:main}, there exists a finite
 stable splitting $\cf$ for $\phi$ and its subformulae such that
 $\cf$ contains $X$ and $U$. Assume that $x, U\sat_\cb \Box \psi$, and
 $V\in\ct$ such that $V\subseteq U$. By Proposition~\ref{prop:part}(b),
 there is some $V' \subseteq U$ in $\cf$ with $V \in \Rem{\cf}{V'}$. By
 Proposition~\ref{prop:basis}, let $W\in\cb$ be such that
 $W\in\Rem{\cf}{V'}$ with $x\in W$. So $x, W\sat_\cb \psi$,
 and thus by induction hypothesis, $x, W \sat_\ct \psi$. By stability,
 twice, $x, V \sat_\ct \psi$ as well.
 \end{proof}

 We are now going to prove that a model based on a topological space
 $\ct$ is equivalent to the one induced by any basis of $\ct$ which is
 lattice. Observe that this enables us to reduce the theory of
 topological spaces to that of spatial lattices and, therefore, to
 answer the conjecture of (Moss and Parikh~\cite{MP}) : a completeness theorem
 for subset spaces which are lattices will extend to the smaller class of
 topological spaces.

 \begin{theorem}
     Let $(X,\ct)$ be a topological space and $\cb$ a basis for $\ct$
 closed under finite unions. Let $\cm_1=\langle X,\ct,i\/\rangle$ and
 $\cm_2=\langle X,\cb,i\/\rangle$ be the corresponding models. Then, for
 all $\phi$, $$\cm_1\sat\phi\quad\Eqv\quad\cm_2\sat\phi.$$
 \label{thm:basis}
 \end{theorem}

 \begin{proof}
 It suffices to prove that $x,U\tsat\phi$ for some $U\in\ct$,
 if and only if $x,U'\bsat\phi$ for some $U'\in\cb$.

 Suppose $x,U\tsat\phi$, where $U\in\ct$, then, by
 Corollary~\ref{corol:subsatm},
 there exists $U'\in\cb$ such that $x\in U'$ and $x,U\tsat\phi$.
 By Corollary~\ref{corol:subsatm}, $x,U'\bsat\phi$.

 Suppose $x,U\bsat\phi$, where $U\in\cb$, then $x,U\tsat\phi$, by
 Corollary~\ref{corol:subsatm}.
 \end{proof}


 \section{Finite Satisfiability}

  \begin{proposition} Let $\langle X,\ct\/\rangle$ be a subset space.
 Let $\cf$ be a finite stable splitting for a formula $\phi$
  and all its subformulae, and assume that $X\in \cf$.
  Then for all $U\in\cf$,
  all $x\in U$, and all subformulae $\psi$ of $\phi$,
  $x, U \sat_\ct \psi$ iff $x, U \sat_\cf \psi$.
 \label{prop:equivalent-models}
  \end{proposition}

  \begin{proof} The argument is by induction on $\phi$.
  The only interesting case
  to consider is when $\phi = \Box\psi$.

  Suppose first that $x, U \sat_\cf \Box\psi$ with $U\in \cf$. We must show
  that $x, U \sat_\ct \Box\psi$ also. Let $V\in\ct$ such that
 $V\subseteq U$; we must show that $x, V\sat_\ct \psi$.
  By Proposition~\ref{prop:part}(b),
  there is some $V' \subseteq U$ in $\cf$ with
  $V \in \Rem{\cf}{V'}$.
  So $x, V' \sat_\cf \psi$, and by
  induction hypothesis, $x, V' \sat_\ct \psi$.
  By stability, $x, V \sat_\ct \psi$ also.

  The other direction (if
  $x, U \sat_\ct \Box\psi$, then
  $x, U \sat_\cf \Box\psi$), is an easy application
  of the induction hypothesis.
 \end{proof}

 Constructing the quotient of $\ct$ under $\sim_{\cf}$ is not adequate
 for generating a finite model because there may still be an infinite
 number of points. It turns out that we only need a finite number of
 them.

 Let $\cm=\langle X,\ct,i\/\rangle$ be a
 topological model, and define an equivalence relation $\sim$ on $X$ by
 $x\sim y$ iff

 \begin{description}
 \item{(a)} for all $U \in \ct$, $x\in U$ iff $y \in U$, and
 \item{(b)} for all atomic $A$, $x\in i(A)$ iff $y \in i(A)$.
 \end{description}

 Further, denote by $x^*$ the equivalence class of $x$,
 and let $X^* = \{x^* \mid  x \in X\}$. For every $U\in\ct$ let  $U^*=\{x^*
 \mid  x\in X\}$, then $\ct^* = \{ U^* \mid  U \in \ct \}$ is a topology on
 $X^*$. Define a map $i^*$ from the atomic formulae to the powerset
 of $X^*$ by $ i^*(A) = \{ x^* \mid  x\in i(A) \} $. The entire model $\cm$
 lifts to the model $\cm^* = \langle X^*,\ct^*, i^* \/\rangle $ in a
 well-defined way.

 \begin{lemma} For all $x$, $U$, and $\phi$,
 $$x, U \sat_\cm \phi \qquad\mbox{iff} \qquad x^*, U^* \sat_{\cm^*} \phi\ .$$
 \label{lemma:quotient}
 \end{lemma}

 \begin{proof}
 By induction on $\phi$.
 \end{proof}

 \begin{theorem}
 If $\phi$ is satisfied in any topological  space then $\phi$
 is satisfied in a finite topological space. \label{thm:finiteness}

 \end{theorem}

 \begin{proof}
 Let $\cm=\langle X,\ct,i\/\rangle$ be such that for some $x\in U \in
 \ct$, $x, U \sat_\cm \phi$.
 Let $\cf^{\phi}$ be a finite stable splitting (by
 Theorem~\ref{thm:main})
 for $\phi$ and its subformulae with respect to $\cm$.
 By Proposition~\ref{prop:equivalent-models},
 $x, U \sat_\cn \phi$, where $\cn=\langle X,\cf,i\/\rangle$. We may
 assume that $\cf$ is a topology,
 and we may also assume that the overall language has only the
 (finitely many) atomic symbols which occur in $\phi$.
 Then the relation $\sim$ has only finitely many classes.
 So the model $\cn^*$ is finite. Finally, by
 Lemma~\ref{lemma:quotient}, $x^*, U^* \sat_{\cn^*}\phi$.
 \end{proof}

 Observe that  the finite
 topological space
 is a quotient of the initial one under two equivalences. The one equivalence is
 $\sim_\cf^\phi$
 on the open subsets of the topological space, where $\cf^\phi$ is the
 finite splitting corresponding to $\phi$ and its cardinality is a
 function of the complexity of $\phi$. The other
 equivalence is $\sim_X$ on the points of the topological space and
 its number of equivalence classes is a function of the atomic formulae
 appearing in $\phi$.
 The following simple example shows how a topology is formed with the quotient
 under these two equivalences

 \begin{example}
 Let $X$ be the interval $[0,1)$ of real line with the
 the set
 $$\ct\quad=\quad\{\emptyset\}\cup\{\ [0,\frac{1}{2^n}) \ \mid \
 n=0,1,2,\ldots\ \}$$
 as topology. Suppose that we have only one atomic formula, call it $A$, such
 that $i(A)=\{0\}$, then it is easy to see that the model $\langle
 X,\ct,i\rangle$ is equivalent
 to the finite topological model $\langle X^*,\ct^*,i^*\rangle$, where
 $$
 \begin{array}{rcl}
 X^* & = & \{\  x_1,x_2\ \},\\
 \ct^* & = & \{\ \emptyset, \{x_1,x_2\}\ \}, \hbox{ and}\\
 i(A) & = & \{\  x_1 \ \}.
 \end{array}
 $$
 \end{example}

   So the overall size of the (finite) topological
 space is bounded by a function of the complexity of $\phi$. Thus if we
 want to test if a given formula is invalid we have a finite number of
 finite topological spaces where we have to test its validity. Thus we
 have the following

 \begin{theorem}
 The theory of topological spaces is decidable.
 \end{theorem}

 Observe that the last two results apply for lattices of subsets by
 Theorem~{\ref{thm:basis}}.

 \medskip

 \noindent{\bf Acknowledgements:} I wish to thank Larry Moss
 and Rohit Parikh for helpful comments and suggestions.


\begin{thebibliography}{1}

 \bibitem[1991]{FHV}
 Fagin,~R., J.~Y. Halpern, and M.~Y. Vardi.
 \newblock A model-theoretic analysis of knowledge.
 \newblock {\em Journal of the Association for Computing Machinery},
   38(2):382--428, 1991.

 \bibitem[1993]{KGT}
 Georgatos,~K.
 \newblock Modal logics for topological spaces.
 \newblock Ph.D. Dissertation. City University of New York, 1993.

 \bibitem[1980]{COMP}
 Gierz,~G., K.~H. Hoffman, K.~Keimel, J.~D. Lawson, M.~W. Mislove, and D.~S.
   Scott.
 \newblock {\em A Compendium of Continuous Lattices}.
 \newblock Springer-Verlag, Berlin, Heidelberg, 1980.

 \bibitem[1984]{HAM}
 Halpern,~J.~Y.,  and Y.~Moses.
 \newblock Knowledge and common knowledge in a distributed environment.
 \newblock In {\em Proceedings of the Third ACM Symposium on Principles of
   Distributed Computing}, pages 50--61, 1984.

 \bibitem[1962]{HI}
 Hintikka,~J.
 \newblock {\em Knowledge and Belief}.
 \newblock Cornell University Press, Ithaca, New York, 1962.

 \bibitem[1992]{MP}
 Moss,~L.~S., and R.~Parikh.
 \newblock Topological reasoning and the logic of knowledge.
 \newblock In Y.~Moses, editor, {\em Proceedings of the Fourth Conference (TARK
   1992)}, pages 95--105, 1992.

 \bibitem[1985]{PR}
 Parikh,~R., and R.~Ramanujam.
 \newblock Distributed computing and the logic of knowledge.
 \newblock In R.~Parikh, editor, {\em Logics of Programs}, number 193 in Lecture
   Notes in Computer Science, pages 256--268, Berlin, New York, 1985.
   Springer-Verlag.

 \bibitem[1968]{RS}
 Rasiowa,~H., and R.~Sikorski.
 \newblock {\em The Mathematics of Metamathematics}.
 \newblock Panstwowe Wydawnictwo Naukowe, Warszawa, Poland, second edition,
   1968.

 \bibitem[1989]{V}
 Vickers,~S.
 \newblock {\em Topology via Logic}.
 \newblock Cambridge Studies in Advanced Computer Science. Cambridge University
   Press, Cambridge, 1989.

 \end{thebibliography}

 \end{document}